\documentclass[aps,showpacs]{revtex4}

\usepackage{amssymb,amsthm}
\newcommand\be{\begin{eqnarray}}
\newcommand\ee{\end{eqnarray}}

\begin{document}

\title{Remarks on Barbero-Immirzi parameter as a field}
  
\date{April 2009}

\author{Alexander Torres-Gomez and Kirill Krasnov}
\affiliation{School of Mathematical Sciences, University of Nottingham, Nottingham, NG7 2RD, UK}

\begin{abstract} We revisit a propagating torsion gravity theory obtained by introducing a
field coupled to the Holst term in the first-order Einstein-Cartan action. 
The resulting theory has second order field equations, no adjustable coupling constants, and one more propagating
degree of freedom as compared to general relativity. When no fermions are present the 
theory is known to be equivalent to that of a single massless scalar field canonically coupled to 
gravity. We extend this result to the case with fermions and obtain an effective interaction between 
the scalar field and the fermionic currents. We also describe a version of the theory with a potential 
for the scalar field and discuss whether it can be interpreted as the inflaton. 
\end{abstract}


\maketitle

\section{Introduction}

Einstein's general relativity (GR) is the only theory of metrics that leads to
field equations that are not higher than second order in derivatives. Thus, to modify GR 
one either has to add to the theory more fields, or increase the order of field equations. 
An interesting modification scheme that has been considered in the literature is to 
start from GR with an additional topological (i.e. total derivative) term added to the action, 
and then promote the coupling constant in front of this topological term into a field. In
a theory of metrics, such as Einstein's GR, a possible topological term
(second order in the curvature) is the Pontryagin density obtained
by a contraction of Riemann curvature with its Hodge dual. Promoting
the coefficient in front of this term into a field (and possibly adding to the
action a kinetic term for this field) leads to the so-called Chern-Simons
modification of general relativity \cite{Deser:1981wh}. The field equations
of this theory are higher than second order in derivatives, which makes them 
non-trivial to analyze. However, things are simplified by the fact that
Schwarzschild and Robertson-Walker spacetimes are still solutions of
the theory. 

Yet another possibility arises when one works not with metrics (second
order formalism) but with the tetrad and the spin connection (first
order formalism). In this case, there is a term linear in the curvature that can be added to the action
and that does not affect the field equations. This term has played a key role in 
the Holst's covariant version \cite{Holst:1995pc} of the Barbero's Hamiltonian description of GR \cite{Barbero:1994ap}.
The Holst term is not a total derivative, so the reason why it does not affect the 
Euler-Lagrange equations is slightly more subtle. The variation of the Holst term
is non-zero, and there is an additional contribution to both the connection and tetrad field equations.
However, for a generic choice of the parameter in front of the Holst term, the connection equation still 
turns out to imply zero torsion, and on this half-shell the contribution from 
the Holst term to the tetrad equation vanishes (due to the cyclic Bianchi identity).

Having an action for GR modified by a term that does not affect the field equations 
one can perform the trick that leads to Chern-Simons gravity 
and promote the coefficient in front of this term, known in the loop quantum gravity literature as 
the Barbero-Immirzi parameter, into a field. The corresponding theory has been known
for a long time, in fact known even before the Barbero-Immirzi parameter was introduced. We 
were able to trace it to the 1991 book by Castellani, D'Auria and Fr\'e
\cite{Castellani:1991et}, see section I.7.2, but it is possible that it was known even earlier. 
The theory has recently been discussed in \cite{Taveras:2008yf}, with the authors apparently being
unaware of the earlier work. Our interest in the theory was prompted by \cite{Taveras:2008yf}, and
only later we came across the reference \cite{Castellani:1991et}.

One of the most interesting features of this theory is that it does not have
adjustable parameters. However, being essentially a scalar-tensor theory with an
additional propagating degree of freedom, one might suspect that it could be easily ruled
out by the classical gravity tests. Indeed, a well-known scalar-tensor theory of
gravity -- Brans-Dicke theory \cite{Brans:1961sx} -- has an adjustable dimensionless parameter,
the coupling constant $\omega$, on which the Solar system tests put a very strict bound.
The original motivation for our work was to see whether one can rule out the theory 
\cite{Taveras:2008yf} using the classical tests. However, the observation of 
\cite{Castellani:1991et} that this theory is effectively equivalent to that of
a single massless scalar field canonically coupled to gravity reduces this problem
to a well-studied one in the context of scalar-tensor theories of gravity, see e.g. 
\cite{Damour:2007uf} for a recent review. One finds that the theory is perfectly consistent 
with all classical, and some non-classical tests, see below, but these results follow 
immediately from the observation of \cite{Castellani:1991et} and are thus not new. 

One novel contribution of this work is a description of the effective theory in
the case fermionic matter is present. As is well-known, in Einstein-Cartan theory 
fermions couple directly to the spin connection, and this coupling introduces 
a torsion into the theory. The theory with a constant Barbero-Immirzi parameter 
coupled to fermions has been analyzed in \cite{Perez:2005pm} and then in 
\cite{Freidel:2005sn}. We extend this analysis to the case of the Barbero-Immirzi 
parameter being a field and obtain the effective coupling of the arising scalar
degree of freedom to fermions. In the case of minimal coupling the effective interaction
term is found to couple the scalar field to the axial fermionic current.

Another contribution of this work is a proposal for a version of the theory with a
natural potential term for the effective scalar field added the action. 
As we shall describe below, a natural potential is characterized by
two parameters. At the level of the arising effective theory these
two parameters translate into the cosmological constant and the 
value to which the Barbero-Immirzi parameter is driven dynamically.
The mass of the scalar propagating degree of freedom is then 
a simple function of these two parameters. We also discuss a
possibility to interpret the scalar field as the inflaton,
which was also the original motivation of \cite{Taveras:2008yf}.
However, our conclusion is that the potential that naturally arises in
this context is too steep to produce any realistic inflationary scenario.

The organization of this paper is as follows. In the next Section we
describe the theory that is our subject of study. Section \ref{sec:eff}
describes the effective second-order theory obtained by integrating
out the spin connection. The new result here is a description of
the coupling between the scalar field and the fermions. Section \ref{sec:tests}
briefly discusses the question whether the theory is consistent with known
gravity tests. Section \ref{sec:pot} introduces and discusses
a version of the theory in which a potential for the scalar field
is added to the action. We conclude with a brief discussion.

\section{The theory}
\label{sec:theory}

Following \cite{Castellani:1991et}, \cite{Taveras:2008yf} we consider a theory of 
gravity described by the following action:
\be\label{action}
S_{gr}[e,\omega,\gamma] = \frac{1}{4} \int_M \epsilon_{IJKL} e^I \wedge e^J \wedge F^{KL} 
+ \frac{1}{2}\int_M \gamma \, e^{I} \wedge e^{J} \wedge F_{IJ}. 
\ee
The quantity $F^{IJ}=d\omega^{IJ} + \omega^I_{\,K} \wedge \omega^{KJ}$ 
is the curvature of the spin connection $\omega^{IJ}$, $e^I$ are the tetrad
one-forms, $\gamma$ is a new dynamical field, and we work in the units $8\pi G=1$. The 
quantity $\epsilon_{IJKL}$ is the totally anti-symmetric tensor in the ``internal'' Minkowski 
space. We use the signature $-,+,+,+$ and $\epsilon_{0123}=1$. With our conventions
we have: $e^I\wedge e^J\wedge e^k\wedge e^L=- \sqrt{-g}\, \epsilon^{IJKL} d^4x$.

The above action describes a ``pure gravity'' theory. To complete the definition
one needs to specify how matter couples to it. Bosonic matter fields (e.g. scalar or gauge) can be
naturally coupled to the metric defined by the tetrad $e^I$. However, in case there are
fermions a choice is to be made. One possibility is to couple fermions
to the spin connection compatible with the metric defined by $e^I$. A more natural possibility,
however, is to couple fermions directly to the connection $\omega^{IJ}$. Here we consider
a single massless Dirac fermion $\psi$ with the action given by:
\be\label{action-ferm}
S_f[e,\omega,\psi] = \frac{i}{12} \int_M \epsilon_{IJKL} e^I\wedge e^J \wedge e^K \wedge
\left( (1-i\alpha) \overline{\psi} \gamma^L D\psi - (1+i\alpha) \overline{D\psi} \gamma^L \psi\right).
\ee
Following \cite{Freidel:2005sn} we have allowed for a non-minimal coupling and $\alpha$
is the associated parameter. The fermion couples directly to the spin connection $\omega^{IJ}$ via
the covariant derivative: $D\psi := d\psi + (1/8) \omega_{IJ} [\gamma^I, \gamma^J] \psi$,
where $\gamma^I$ are the Dirac matrices. Additional fermions as well as a mass term for them can 
be easily added but do not change the analysis that follows. 

As is pointed out already in \cite{Castellani:1991et}, the transformation properties of the Einstein-Cartan
and the Holst term under spacetime reflections are different: the former transforms as
a scalar, the latter as a pseudoscalar. Thus, if, as is suggested in \cite{Castellani:1991et},
the new scalar field is chosen to transform as a pseudoscalar, the resulting theory is parity invariant. 
The choice $\gamma$ being a pseudoscalar is also supported by the fact that the resulting
torsion components have well-defined parity transformation properties in this case, 
see (\ref{T-S}). This is also the standard choice in other cases where a coefficient
in front of a topological term is promoted into a field, e.g. the so-called
axion electrodynamics, see \cite{Wilczek:1987mv}. So, in this paper we shall
think about $\gamma$ as a pseudoscalar as well. In this case the limit to the
Barbero-Immirzi parameter case $\gamma=const$ is quite peculiar, for this
constant now is parity odd. Note, however, that this is not in conflict with the loop quantum
gravity applications, where the Barbero-Immirzi parameter $\gamma=const$
appears as a prefactor in the spectra of geometric operators such as areaa
and volume. Indeed, the area and volume then change the sign when
the spatial orientation is reversed, but this is as expected from such geometric 
quantities~\footnote{We thank an anonymous referee for pointing out this interpretation 
of the pseudoscalar Barbero-Immirzi parameter $\gamma$ to us.}.

A related issue is that of $\gamma$-field reality conditions. The 
first attitude might be that $\gamma$ must be real in order for the Lagrangian to be real. 
However, let us consider a single particle described by its position
$q$ and momentum $p$ with the action $S=\int dt(p \dot{q} - i\lambda p - \lambda^2/2)$.
The Lagrangian is not real. However, one gets the usual real free particle 
Lagrangian on-shell $\lambda=p/i$. We learn that, provided some of the variables in
the action are ``auxiliary'', it is possible to have a non-real Lagrangian 
without any physical inconsistencies. One must only require that after all such auxiliary variables
are eliminated the arising effective Lagrangian is real. Below we shall see that in
the case of theory (\ref{action}) the spin connection $\omega^{IJ}$ is such an auxiliary variable,
and that, in the case of a ``pure gravity'' theory without fermions both $\gamma$ real and purely imaginary 
lead to real Lagrangians. For readers familiar with Plebanski's self-dual formulation \cite{Plebanski} of GR
the possibility of imaginary values of $\gamma$ in (\ref{action}) should not come as a surprise. Indeed,
this formulation corresponds to $\gamma=\pm i$. 

However, the physical consistency of a theory
with a complex Lagrangian cannot be ascertained without specifying couplings to other
fields. A result relevant to our case is that of  \cite{Freidel:2005sn} where it was shown
that imaginary constant values of $\gamma$ lead to a complex effective Lagrangian when 
the non-minimal coupling parameter $\alpha$ is different from zero. Thus, if one is to insist on such
a coupling, then the only consistent choice is $\gamma$ real. Moreover, as we
shall see in the next section, when $\gamma\not= const$ one gets an additional
term coupling the $\partial_\mu \gamma$ and fermionic currents. This term is imaginary when 
$\gamma$ is imaginary even in the case of minimal coupling. Thus, if one insists that the fermionic
matter couples directly to the spin connection $\omega$, 
then the only option that leads to a real effective Lagrangian is to require $\gamma$ to be real.
Importantly, as we shall see below, this is also the case in which the effective scalar field carries
positive energy.

The Euler-Lagrange equations obtained from the action (\ref{action}) are as follows:
\be\label{eq-comp}
{\cal D}( \epsilon_{IJKL} e^K \wedge e^L + 2 \gamma e_I \wedge e_J)=0, \\
\label{eq-gamma}
e^{I} \wedge e^{J} \wedge F_{IJ}=0, \\
\label{eq-tetrad}
\epsilon_{IJKL} e^J \wedge F^{KL} + 2 \gamma e^J \wedge F_{IJ} = 0.
\ee
When matter is present the first and last of these equations get matter contributions.
The first equation only gets a contribution from the fermionic term.
The equation (\ref{eq-comp}) is an algebraic equation for the components of the spin connection 
$\omega^{IJ}$. It can be solved for $\omega^{IJ}$ in terms of the derivatives of the tetrads and the 
field $\gamma$ (and fermionic currents in case when fermions are present). This solution can 
be substituted into (\ref{eq-gamma}) to obtain a dynamical equation for the $\gamma$-field,
and into (\ref{eq-tetrad}) to obtain equations for the components of the tetrad. Alternatively,
one can substitute the solution for $\omega^{IJ}$ into the action (\ref{action}), and
obtain an effective theory. 

\section{The effective second-order theory}
\label{sec:eff}

In this section we solve the equation for the spin connection and use the solution
obtained to exhibit an effective second-order theory that involves only the metric,
not the spin connection, as an independent variable. In the case without fermions this exercise has
been carried out in \cite{Castellani:1991et} and more recently in \cite{Taveras:2008yf}.
The case $\gamma=const$ has been analyzed in detail in \cite{Freidel:2005sn}. Here
we extend the analysis to the case $\gamma\not= const$ and fermions being present.

When fermionic fields are present the right-hand-side of (\ref{eq-comp}) receives
an additional contribution from the fermionic part of the action. It has been
worked out in \cite{Freidel:2005sn} and is unchanged. However, the authors have
used the component notation. Here is the same derivation using the form notation. 
The variation of the fermionic action with respect to the connection one-form is:
\be\label{f-var}
\frac{\delta S_f}{\delta \omega^{IJ}} = \frac{i}{12*8} \epsilon_{KLMN} e^K \wedge e^L \wedge e^M
\left( \overline{\psi} \{\gamma^N,[\gamma_I,\gamma_J]\} \psi 
- i\alpha \overline{\psi} [\gamma^N,[\gamma_I,\gamma_J]] \psi\right),
\ee
where $\{\cdot,\cdot\}$ is the anti-commutator. Using the identities 
$[\gamma^N,[\gamma^I,\gamma^J]]=8\delta^N_{[I} \gamma_{J]}$
and $\{\gamma^N,[\gamma_I,\gamma_J]\}=-4i \epsilon^{N}_{\, IJK} \gamma^5 \gamma^K$
(this identity holds with our choices for the signature and $\epsilon_{0123}=1$ if
one defines $\gamma^5=(1/i)\gamma^0\gamma^1\gamma^2\gamma^3$) 
and introducing the axial and ordinary vector fermionic currents:
\be
A^I = \overline{\psi} \gamma^5 \gamma^I \psi, \qquad V^I = \overline{\psi} \gamma^I \psi,
\ee
we can write (\ref{f-var}) as:
\be
\frac{\delta S_f}{\delta \omega^{IJ}} = \frac{1}{24} \epsilon_{KLMN} e^K \wedge e^L \wedge e^M \left(
\epsilon^N_{\,IJP} A^P + 2\alpha \delta^N_{[I} V_{J]}\right).
\ee
Thus, the equation (\ref{eq-comp}) in the presence of fermions becomes:
\be
{\cal D}( \epsilon_{IJKL} e^K \wedge e^L + 2 \gamma e_I \wedge e_J)=4 \frac{\delta S_f}{\delta \omega^{IJ}} 
\ee
This equation can be viewed as an algebraic equation for the components of the spin connection. 
A particularly efficient way to solve it is to introduce the metric-compatible connection
$\Gamma^{IJ}: D^{\Gamma} e^I=0$. Then $\omega^{IJ} = \Gamma^{IJ} + C^{IJ}$, where
$C^{IJ}$ are the contorsion one-forms. Introducing a convenient notation:
\be
B^{IJ} = \frac{1}{2} \epsilon^{IJKL} e_K \wedge e_L + \gamma e_I \wedge e_J
\ee
we can write the resulting equation for the coefficients $C^{IJ}$ as:
\be
2 C^{[I}_{\,\,\,\,K} \wedge B^{|K| J]} + d\gamma \wedge e^I\wedge e^J = 2 \frac{\delta S_f}{\delta \omega^{IJ}}.
\ee
Its solution in the absence of fermions was found in \cite{Castellani:1991et} and \cite{Taveras:2008yf}.
The solution in the case $\gamma=const$ and fermions present was obtained in \cite{Freidel:2005sn}.
Since this is a linear equation on $C^{IJ}$ its solution is given by the sum of these
two known solutions. In our notations, we get:
\be\label{torsion}
C^{IJ} = - \frac{1}{2(\gamma^2+1)} 
\left( \epsilon^{IJKL} e_K \partial_L \gamma - 2\gamma e^{[I} \partial^{J]}\gamma \right)
\\ \nonumber
+ \frac{1}{4(\gamma^2+1)} \left( \epsilon^{IJKL} e_K (A_L + \alpha\gamma V_L) -
2e^{[I} (\gamma A^{J]} - \alpha V^{J]}) \right).
\ee 

It is also illuminating to write down an explicit expression for the torsion tensor. 
The relation between the torsion 2-form and the contorsion 1-form is $T^I=C^I\,_J \wedge e^J$. 
Thus, we get:
\be
T^{I}_{\mu \nu}=\frac{1}{\gamma^2 +1} 
\left(  \epsilon^I_{\mu \nu \lambda} 
\left(  \partial^{\lambda}\gamma -\frac{1}{2} A^{\lambda}-\frac{\alpha \beta}{2} V^{\lambda}  \right) 
+ \delta^I_{[\mu} \left( \gamma \, \partial_{\nu]}\gamma -\frac{\gamma}{2} A_{\nu]} +\frac{\alpha}{2} V_{\nu]}   
\right)      \right) \, ,
\ee
where the fermionic currents are $A_\mu = e_\mu^I A_I, V_\mu=e_\mu^I V_I$.
The torsion tensor $T^\sigma_{\mu\nu}:=e^\sigma_I T^I_{\mu\nu}$ can then be decomposed into three 
irreducible components with respect to the Lorentz group, which are the trace vector 
$T_{\mu}=e_I^{\nu}\, T^{I}_{\mu \nu}$, the axial vector 
$S_{\mu}= \epsilon_{\mu \nu \rho \sigma} T^{\nu \rho \sigma}$ and a tensorial component. 
In our case the trace and the axial vector components are given by:
\be\label{T-S}
T_{\mu}=-\frac{3}{2(\gamma^2 +1)} 
\left( \gamma \, \partial_{\mu} \gamma - \frac{\gamma}{2} A_{\mu}+\frac{\alpha}{2} V_{\mu}  \right) \, ,
\\ \nonumber
S_{\mu}=-\frac{6}{\gamma^2 +1} 
\left( \partial_{\mu} \gamma - \frac{1}{2} A_{\mu}-\frac{\alpha \, \gamma}{2} V_{\mu}  \right) \, ,
\ee
respectively. It is clear that, if $\gamma$ is an ordinary parity even scalar, then these two vectors do not
have simple transformation properties under parity, a point emphasized in particular in
\cite{Mercuri:2006um}. This is of no surprise, since in this case the 
action contains a parity-violating term. If, on the other hand, one chooses the field
$\gamma$ to be a pseudo-scalar, then the torsion components $T_\mu, S_\mu$ transform as a vector
and pseudovector correspondingly. This motivation for the choice $\gamma$ being a pseudoscalar
is analogous to the reasonigs that leads to the conclusion that the axion is a pseudoscalar
in the axion electrodynamics \cite{Wilczek:1987mv}. In the case of electrodynamics the
pseudoscalar nature of the axion follows from the requirement that the electric and
magnetic fields have the usual transformation properties under the space reversals, in the
case under consideration the same role is played by the torsion components. 

To obtain an effective theory one simply has to substitute the resulting connection into
the actions (\ref{action}), (\ref{action-ferm}). One uses
$F^{IJ} = R^{IJ} + D^{\Gamma} C^{IJ} + C^{I}_K\wedge C^{KJ}$. The part of the result independent
of the fermions has been worked out in \cite{Castellani:1991et} and \cite{Taveras:2008yf}
and is given by:
\be\label{eff-gr}
S^{eff}_{gr}= \frac{1}{2} \int_M \sqrt{-g} 
\left( R - \frac{3}{2(\gamma^2+1)} g^{\mu\nu} \partial_\mu \gamma \partial_\nu \gamma\right).
\ee
Our new result is that for a part of the effective action that couples the scalar field
$\partial_\mu \gamma$ and fermionic currents. One finds that all contributions to this
from (\ref{action}) cancel, so it remains to compute only the contribution from (\ref{action-ferm}).
This is an easy exercise with the following result:
\be\label{eff-g-ferm}
S^{eff}_{f\gamma} = \frac{1}{2} \int_M \sqrt{-g} \left(
\frac{3}{2(\gamma^2+1)} g^{\mu\nu} \partial_\mu \gamma ( A_\nu + \alpha \gamma  V_\nu)\right).
\ee
This is a simple coupling between the $\gamma$ and fermionic currents, and, in fact,
its structure could have been expected. Finally, the part of the effective action that does not 
contain derivatives of $\gamma$ is:
\be\label{eff-ferm}
S^{eff}_f=
\frac{i}{2} \int_M \sqrt{-g} \left( (1-i\alpha) \overline{\psi} \gamma^\mu D^\Gamma_{\mu} \psi 
- (1+i\alpha) \overline{D^\Gamma_{\mu} \psi} \gamma^\mu \psi\right)
\\ \nonumber + 
\frac{1}{2} \int_M \sqrt{-g} \left( \frac{3}{8(\gamma^2+1)} (A^2 + 2\alpha\gamma AV - \alpha^2 V^2)\right),
\ee
which agrees with \cite{Freidel:2005sn}.

As has been observed in \cite{Castellani:1991et} (and apparently
unnoticed in \cite{Taveras:2008yf}\footnote{The authors of \cite{Taveras:2008yf} inform us
that this fact was known to them and, in particular, stated in their seminars, but not mentioned in the paper.}), 
after a change of variables 
\be
\gamma=\sinh(\chi),
\ee
where $\chi$ is a new field, the action (\ref{eff-gr}) of the ``pure'' effective scalar-tensor 
theory takes the form:
\be\label{action-eff}
S[g,\chi]=\int_M \sqrt{-g} 
\left( R - \frac{3}{2} g^{\mu\nu} \partial_\mu \chi \partial_\nu \chi\right).
\ee
However, as is clear from (\ref{eff-g-ferm}) and (\ref{eff-ferm}), the coupling of the
effective field $\chi$ to fermions is quite non-standard, with an additional prefactor
of $1/\cosh(\chi)$ being present in front of the canonical $\chi$-$\psi$ current interaction terms.

\section{Gravity tests}
\label{sec:tests}

As we have seen in the previous Section, in the absence of fermions the theory becomes
equivalent to that of a massless scalar field $\chi$ canonically coupled to gravity.
Thus, to see whether this theory is consistent with observations one can use available
results on scalar-tensor theories. These results, see e.g. \cite{Damour:2007uf}
for a recent review, immediately imply that the theory in question not only passes the classical gravity 
tests, but is also compatible with binary pulsar observations. Indeed, in the classification
scheme of this reference, the theory (\ref{action-eff}) is one in which the logarithmic coupling 
function $a(\phi)$ vanishes. This means that for all the tests discussed in
\cite{Damour:2007uf}, i.e. the solar system as well as the binary pulsar tests, 
the theory (\ref{action-eff}) is indistinguishable from GR. 

The only remaining standard gravity test is by the cosmological scenario that it produces. In
\cite{Taveras:2008yf} it was noted that the homogeneous isotropic cosmology of the theory
in question is that of a $\rho=P$ ideal fluid. This fact again easily follows 
from the effective scalar-tensor description. The energy density of such a fluid behaves
as $a^{-6}$, where $a$ is the scale factor of the Universe. Therefore, such a fluid only
matters in the very early stages of the history of the Universe and would not have any effect on the structure
formation and the arising CMB spectrum. Indeed, the phenomena leading to both occur much later -- 
during the transition from the radiation to matter domination. Thus, it seems that the modification of 
gravity in question is also innocuous for cosmology. We conclude that the described simple 
generalization of the first-order Einstein-Cartan theory gives rise to a viable gravity theory.

Being a scalar-tensor theory (\ref{action-eff}) is related to Brans-Dicke theory
\cite{Brans:1961sx} by a simple conformal transformation. Thus, introducing a new
scalar field $\phi=\exp{\chi}$ and a new metric $g_{BD}=\exp(-\chi) g$, we obtain
the $\omega=0$ Brans-Dicke theory for $g_{BD}, \phi$. Have matter fields coupled
to $g_{BD}$ the theory would be in violation of the solar system tests which require
the Brans-Dicke parameter to be $\omega > 10^5$, see e.g. \cite{Damour:2007uf}. However, 
the situation is different in our case with matter coupling to $\phi\, g_{BD}$. 
For completeness of our exposition of this theory, in appendix we review how the weak field 
expansion of its spherically-symmetic solution is consistent with the Solar System tests.

\section{Version with a potential}
\label{sec:pot}

In the previous two Sections we have seen that the theory is equivalent to that of
a single massless scalar field and that it is perfectly viable as a gravity theory. Thus, 
if such a varying $\gamma$ (or $\chi$) field existed, in particular, in the Solar system we would
not be aware of it. However, general relativity can achieve the same consistency with observations 
without any additional scalar fields, so why consider any alternative such as (\ref{action})? 

Recall that in a homogeneous isotropic Universe our massless scalar field behaves as a $\rho=P$
ideal fluid and is thus only important during the very early stages of the expansion. This suggests
that it may be possible to utilize this field in a mechanism at play during the early stages of
the history of the Universe. In particular, the natural question, which was also the original motivation of
the paper \cite{Taveras:2008yf}, is whether one can interpret $\gamma$ (or $\chi$) as the inflaton.

This is obviously not possible with the theory in its version (\ref{action}) as it gives rise
to a massless effective field. However, as we shall now see, a certain natural potential for the 
$\gamma$ (and thus $\chi$) field can be added to the action. Thus, let us note that the action (\ref{action}) 
is that of $BF$ theory with the $B$-field given by:
\be\label{B}
B^{IJ} = \frac{1}{2} \epsilon^{IJKL} e_K \wedge e_L + \gamma e^I \wedge e^J.
\ee
Thus, one can naturally add to the action two ``cosmological constant'' terms quadratic in $B^{IJ}$.
Let us first consider the more physical case of real $\gamma$. Then the terms to be added are:
\be\label{terms}
-\frac{a}{2*4!} \epsilon_{IJKL} B^{IJ} \wedge B^{KL} = -\frac{a}{2}( \gamma^2 -1) \, \sqrt{-g} d^4 x, \\ \nonumber
-\frac{b}{4!} B_{IJ} \wedge B^{IJ} = -b\gamma  \, \sqrt{-g} d^4 x.
\ee
The action (\ref{action}) with these terms added gives an effective scalar-tensor theory for the
field $\chi$ with the potential:
\be\label{pot}
V(\chi) = \frac{a}{2}(\sinh^2(\chi)-1) + b\sinh(\chi).
\ee
In order for this potential to have a minimum (so that the associated excitations have positive mass) 
the parameter $a$ has to be chosen to be positive. Then the minimum occurs at $\gamma_m = \sinh(\chi_m)=-b/a$. 
The value of the potential at the minimum is $V(\chi_m)=-(a/2)(1+b^2/a^2)$ and this
should be interpreted as the cosmological constant of the theory $\Lambda=V(\chi_m)$. Note that it is negative. 
Thus, the theory can naturally describe a negative cosmological constant. The mass of the excitations
around the minimum is $m^2_\chi=a(1+b^2/a^2)=2|\Lambda|$. Thus, the two parameters characterizing
the version of the theory with the potential can be taken to be: a negative cosmological 
constant $\Lambda$ and the value $\gamma_m$ of the field at the minimum of the potential.

We see that the natural potential for the $\gamma$-field that can be added to the
action does not seem to be physically realistic as it gives a negative cosmological constant.
It is easy to rectify this by considering the case of the purely imaginary field $\gamma$.
However, this is not a viable scenario since the effective field has negative kinetic energy in this
case, and the theory is inconsistent (has complex Lagrangian) when it is coupled
to fermions directly via the spin connection. 

One may now wonder if it is possible to get an inflating Universe out of the model
with potential (\ref{pot}). Unfortunately, as simple considerations show, this is {\it not} a 
model for inflation. Indeed, to inflate the Universe has to start
at values of the field sufficiently far away from the minimum of the potential. In
that regime one can approximate the potential by an exponent $V\sim \exp(2\chi)$. 
However, it is well known, see e.g. \cite{Ratra:1989uz}, that to get a realistic expansion in a 
model with exponential potential the coefficient in the exponent should be much smaller than one. 
The potential $V\sim \exp(2\chi)$ is thus obviously too steep to serve as an inflationary one.
Another way to arrive at the same conclusion is to note that the standard slow roll
conditions $(V_{,\chi}/V)^2 \ll 1, |V_{,\chi\chi}/V| \ll 1$ are obviously violated
by the potential $V\sim \exp(2\chi)$.

Thus, we conclude that, in spite of the fact that a certain natural potential (\ref{pot}) for the field
$\gamma$ can be added to the action (\ref{action}), this potential does not lead to 
any interesting physics. Its minimum, which sets the value of the cosmological constant
in the theory, is negative, and it is too steep to be of any interest for inflationary
model building.

\section{Discussion}

In this short paper we have revisited a certain simple modification of the Einstein-Cartan
formulation of general relativity. This modification has been previously considered in
\cite{Castellani:1991et}, and more recently in \cite{Taveras:2008yf}, and in both references
an effective action for the scalar degree of freedom was obtained. In \cite{Castellani:1991et} 
it was in addition observed that a simple field redefinition renders
the effective action into that of a massless scalar field coupled to gravity. 
Most of the physical properties of the model can thus be obtained from this
equivalence to a simple scalar-tensor gravity theory.

We have discussed in details the issue of reality conditions that need to
be imposed on the field $\gamma$. We have seen that this question is best
settled by considering the coupling of theory to matter, fermions in particular.
We have seen that if one insists on coupling fermions directly to the spin connection,
which seems quite sensible, then the only option that leads to a consistent
theory is to require $\gamma$ to be real. Importantly, in this case the
kinetic energy of the effective scalar field is positive.

We have extended the analysis of the effective second-order theory to the
case when fermions are present and found (in the minimal cooupling case $\alpha=0$) 
an interesting interaction between the $\gamma$-field $\partial_\mu\gamma$ and the fermionic axial
$\overline{\psi} \gamma^5 \gamma^\mu \psi$ currents. If $\gamma$ is a pseudoscalar,
which is a choice preferred by the simple parity transformation properties of the torsion
tensor, then then this term is explicitly parity-invariant. However, if one instead decided to 
define $\gamma$ to be a parity even scalar this term would lead to parity violation even in the 
case of the minimal coupling. 

We have also considered a version of the theory with a potential
for the $\gamma$-field, motivated mainly by the question whether
an inflationary model can be realized in this framework. However,
we have found that a natural potential that can be added
is not of any physical interest - it leads to a negative
cosmological constant and is too steep to produce inflation.

We would like to finish our paper with some remarks. The fact that the same
$w=0$ Brans-Dicke scalar-tensor gravity theory appears in our context (\ref{action}) 
and in the area metric theory studied in \cite{Punzi:2008dv} suggests that one may be 
dealing with a related theory. It would be important to establish this relation, if it exists, 
to decrease the number of different modification schemes available in the literature.

The final point that is worth emphasizing is as follows. In the retrospect,
the theory (\ref{action}) can be seen to be a way to describe an ordinary
scalar field in the framework of theories whose dynamical variables are
spacetime forms. Such formulations of gravitational theories in particular play the key
role in the programme of loop and spin foam quantization of GR. However, it was
always considered to be problematic to incorporate scalar fields into such a framework. 
What the formulation (\ref{action}) does is precisely this - it describes a scalar field
as a part of the two-form $B$-field (\ref{B}) that at the same time encodes 
information about the spacetime metric. The theory then becomes that of BF-type,
and these are accessible to the spin foam quantization methods. 
The price to pay for this ``unification'' of the metric
and the scalar degree of freedom is that the coupling of the scalar field to fermions
is fixed and unconventional, see (\ref{eff-g-ferm}), (\ref{eff-ferm}). We have
also seen that the scalar field potential that can be naturally built from the $B$-field
is not one desired. However, the lessons learned from (\ref{action}) are probably
worth keeping in mind, for it may be that more realistic scenarios for coupling
scalar fields to gravity can be built following an analogous strategy.

{\bf Acknowledgements.} ATG was supported by a Mathematical Science Research Scholarship and KK 
by an EPSRC Advanced Fellowship. The authors thank Jorma
Louko for a discussion, Yuri Shtanov for pointing out reference \cite{Ratra:1989uz}, 
as well as Abhay Ashtekar and the authors of \cite{Taveras:2008yf} for correspondence.

\section*{Appendix: Spherically-symmetric solution}

In this appendix we briefly describe the spherically-symmetric solution of the
gravity theory that is the subject of this paper. Being equivalent to a scalar-tensor
theory, the corresponding solution could be obtained from the known one for
e.g. Brans-Dicke theory \cite{Brans:1962} by a conformal transformation. 
However, we believe it is illuminating to see how this solution arises directly
via the connection and tetrad one-forms. The formulas for the spherically-symmetric 
non-torsion-free spin connection presented below are new, and would be necessary should one
decide to study the physics of test fermions on a spherically-symmetric background with a non-trivial
$\gamma$ profile.

The spherically-symmetric ansatz for the metric is the standard $ds^2 = -f^2 dt^2 + g^2 dr^2 + r^2 d\Omega^2$,
where $f, g$ are unknown functions of the radial coordinate $r$. The field $\gamma$ 
(and thus $\chi$) is also assumed to depend only on the radial coordinate. The spin connection
coefficients are found to be:
\be\label{conn}
w^{0}_{\,\, 1} = \frac{1}{g}\left( f'+ \frac{\gamma \gamma' f}{2\gamma^2+2}\right) dt, \quad
w^{0}_{\,\, 2} = \frac{\sin(\theta)}{g} \left( \frac{\gamma' r}{2\gamma^2 +2} \right) d\phi, \quad
w^{0}_{\,\, 3} = -\frac{1}{g} \left( \frac{\gamma' r}{2\gamma^2 +2} \right) d\theta, \\ \nonumber
w^{1}_{\,\, 2} = - \frac{1}{g} \left( 1 + \frac{\gamma \gamma' r}{2\gamma^2+2}\right) d\theta, \quad
w^{1}_{\,\, 3} = - \frac{\sin(\theta)}{g} \left( 1 + \frac{\gamma \gamma' r}{2\gamma^2+2}\right) d\phi, \quad
w^{2}_{\,\, 3} = - \cos(\theta) d\phi - \frac{\gamma' f}{g(2\gamma^2+2)} dt.
\ee
Substituting these into (\ref{eq-gamma}) one finds the dynamical equation for the field $\gamma$, which,
after the already mentioned field redefinition $\gamma=\sinh(\chi)$ becomes $\Box \chi=0$. This gives:
\be\label{laplace}
\partial_r \left((f/g) r^2 \partial_r \chi \right)= 0,
\ee
whose first integral is:
\be\label{sol-chi}
\chi' = \frac{K g}{r^2 f}.
\ee
Here $K$ is an integration constant that has the dimensions of length. Both real and imaginary
values of this parameter lead to real metrics, but imaginary $K$ corresponds to imaginary $\gamma$.
Our discussion of the coupling to fermions above has demonstrated that for imaginary non-constant
$\gamma$ the theory is inconsistent. Thus, we shall assume $K$ to be real.

To obtain differential equations for the functions $f,g$, we use equations (\ref{eq-tetrad}) and 
substitute the expressions (\ref{conn}) for the spin connection coefficients. The obtained 
equations coincide with Einstein equations $G_{\mu\nu} = T_{\mu\nu}$, with the diagonal 
\be\label{rho}
\rho=P=\frac{3(\chi')^2}{4g^2} = \frac{3K^2}{4 r^4 f^2}
\ee
stress energy tensor that is obtained from the $\chi$-part of the action (\ref{action-eff}).
To write the last equality in (\ref{rho}) we have used (\ref{sol-chi}). 
With our choice of units $8\pi G=1$, the $00,11$ Einstein equations are:
\be
 1-\frac{1}{g^2}+2\,r \frac{g'}{g^3}=r^2 \rho, \qquad 
 -1+\frac{1}{g^2}+\frac{2\,r\,f'}{f g^2}= r^2 P. \label{eqf1}
\ee 
Taking the sum and the difference of these equations we get, after some simple algebra:
\be\label{eq-fg}
\frac{(gf)'}{gf} = \frac{3K^2}{4r^3}(g/f)^2, \qquad 
gf=(rf/g)'.
\ee
Let us now introduce a function $\xi = rf/g$. Then $gf=\xi'$. As we shall see, it is 
this function that will play the role of a convenient radial coordinate for the solution. 
We can now rewrite (\ref{eq-fg}) as
\be
\frac{\xi''}{\xi'} = \frac{3K^2}{4r \xi^2}.
\ee
Integrating this equation once gives:
\be\label{eq-xi}
\xi' = \frac{(\xi+r_+)(\xi+r_-)}{r\xi},
\ee
where $r_\pm = \left( R \pm \sqrt{R^2+3K^2}\right)/2$
and $R$ is a new integration constant of dimensions of length. 
 
Let us now integrate (\ref{eq-xi}). Assuming $r_+\not = r_-$ we get:
\be\label{xi}
(\xi+r_+)^{\frac{r_+}{r_+-r_-}} (\xi+r_-)^{\frac{r_-}{r_- -r_+}}= r.
\ee
Here we have set the arising integrating constant to a particular value so that for 
$r\to\infty, \xi(r)\sim r$. Indeed, for large radii we would like both $f,g$ to approach
unity, and so $\xi=r f/g$ must go like the radial coordinate. 

In terms of $\xi(r)$ the functions $f,g$ are given by:
\be\label{fg}
f^2 = \frac{(\xi+r_+)(\xi+r_-)}{r^2}, \qquad
g^2 = \frac{(\xi+r_+)(\xi+r_-)}{\xi^2}.
\ee
This solves the problem of computing the metric. As we see, the solution is parametrized by two 
constants $R, K$ of dimensions of length.  As we shall soon see, the quantity $R$ is essentially the mass 
of our spherically-symmetric object as seen from infinity $R\sim M$, while $K$ is a parameter that describes 
the modification. The Schwarzschild solution is readily obtained by setting $K=0$. 

To determine the post-Newtonian predictions of the theory we expand the metric 
in powers of $1/r$. An asymptotic relation between the $\xi$ and $r$ coordinates is: 
\be\nonumber
\frac{\xi}{r}=1-\frac{r_++r_-}{r}-\frac{r_+ r_-}{2r^2}- \frac{r_+ r_- (r_+ + r_-)}{3r^3}+ \ldots,
\ee 
from where one easily obtains the asymptotic expansions for the metric functions:
\be\nonumber
f^2=1-\frac{r_+ + r_-}{r}-\frac{r_+ r_- (r_+ + r_-)}{6 r^3}+\ldots, \qquad
g^2=1+\frac{r_+ + r_-}{r}+\frac{r_+ r_- + (r_+ + r_-)^2}{r^2} + \ldots.
\ee
It is useful to rewrite these expansions in terms of the original parameters $R, K$. We get:
\be\label{exp}
f^2=1-\frac{R}{r}+ \frac{RK^2}{8r^3}+ O(1/r^4), \qquad
g^2=1+\frac{R}{r}+\frac{R^2 - 3K^2/4}{r^2} + O(1/r^3).
\ee
We see that the parameter $R$ is essentially the mass of the object, and thus must be positive. 
The most striking features about the expansions (\ref{exp}) are 
that there is no $1/r^2$ term in the expression for $f^2$, and the coefficient in front of $1/r$ term 
in $g^2$ is minus the coefficient in front of the $1/r$ term in $f^2$, which is exactly like in GR. 
We see that the theory (\ref{action}) behaves very much like general relativity, at least in the weak field 
limit. However, in view of its equivalence to (\ref{action-eff}) and the known results on scalar-tensor theories
this is not surprising. 

The solution obtained is that of GR coupled to a massless scalar field. The natural
question that arises is for what values of parameters $R,K$ is this solution a black hole. The
answer to this question is well known \cite{Bekenstein:1971hc}: static black holes 
cannot be endowed with any classical massless or massive scalar field hair. Thus, only the 
$K=0$ Schwarzschild solution with $\gamma=const$, can be a black hole. 
In other words, none of the modified solutions (i.e. solutions with non-trivial profile of the scalar field) 
is a black hole \footnote{This statement was given in a misleading way in the first version of this paper
placed on the arxiv.}. It is also not hard to see this directly from the solution we have desribed.


\begin{thebibliography}{0}

\bibitem{Deser:1981wh}
  S.~Deser, R.~Jackiw and S.~Templeton,
  Annals Phys.\  {\bf 140}, 372 (1982)
  [Erratum-ibid.\  {\bf 185}, 406.1988\ APNYA,281,409 (1988\ APNYA,281,409-449.2000)].

\bibitem{Holst:1995pc}
  S.~Holst,
  Phys.\ Rev.\  D {\bf 53}, 5966 (1996)
  [arXiv:gr-qc/9511026].

\bibitem{Barbero:1994ap}
  J.~F.~Barbero G.,
  Phys.\ Rev.\  D {\bf 51}, 5507 (1995)
  [arXiv:gr-qc/9410014].

\bibitem{Castellani:1991et}
  L.~Castellani, R.~D'Auria and P.~Fre,
{\it  Singapore, Singapore: World Scientific (1991) 1-603}

\bibitem{Taveras:2008yf}
  V.~Taveras and N.~Yunes,
  Phys.\ Rev.\  D {\bf 78}, 064070 (2008)
  [arXiv:0807.2652 [gr-qc]].

\bibitem{Brans:1961sx}
  C.~Brans and R.~H.~Dicke,
  Phys.\ Rev.\  {\bf 124}, 925 (1961).

\bibitem{Damour:2007uf}
  T.~Damour,
  arXiv:0704.0749 [gr-qc].

\bibitem{Perez:2005pm}
  A.~Perez and C.~Rovelli,
  Phys.\ Rev.\  D {\bf 73}, 044013 (2006)
  [arXiv:gr-qc/0505081].

\bibitem{Freidel:2005sn}
  L.~Freidel, D.~Minic and T.~Takeuchi,
  Phys.\ Rev.\  D {\bf 72}, 104002 (2005)
  [arXiv:hep-th/0507253].

\bibitem{Wilczek:1987mv}
  F.~Wilczek,
  Phys.\ Rev.\ Lett.\  {\bf 58}, 1799 (1987).

\bibitem{Plebanski} J. Plebanski, Journ. Math. Phys. {\bf 18},  2511  (1977).

\bibitem{Mercuri:2006um}
  S.~Mercuri,
  Phys.\ Rev.\  D {\bf 73}, 084016 (2006)
  [arXiv:gr-qc/0601013].

\bibitem{Ratra:1989uz}
  B.~Ratra,
  Phys.\ Rev.\  D {\bf 45}, 1913 (1992).

\bibitem{Punzi:2008dv}
  R.~Punzi, F.~P.~Schuller and M.~N.~R.~Wohlfarth,
  arXiv:0804.4067 [gr-qc].

\bibitem{Brans:1962}
  C.~Brans,
  Phys.\ Rev.\  {\bf 125}, 2194 (1962).

\bibitem{Bekenstein:1971hc}
  J.~D.~Bekenstein,
  Phys.\ Rev.\  D {\bf 5}, 1239 (1972).


\end{thebibliography}
\end{document}